\documentstyle[preprint,tighten,aps]{revtex}
\begin{document}           %
\draft
\preprint{\vbox{\noindent
          \null\hfill astro-ph/9509116\\
          \null\hfill LNGS-95/96\\
          \null\hfill  INFNFE-15-95\\
          \null\hfill  INFNCA-TH9511}}
\title{ LAST HOPE \\
for an astrophysical solution to the solar neutrino problem.\\
      }
\author{
         V.~Berezinsky,$^{(1)}$
         G.~Fiorentini$,^{(2)}$
         and M.~Lissia$^{(3)}$
       }
\address{
$^{(1)}$INFN, Laboratori Nazionali del Gran Sasso,
       67010 Assergi (AQ), Italy \\
$^{(2)}$Dipartimento di Fisica dell'Universit\`a di Ferrara,
        I-44100 Ferrara, \\ and Istituto Nazionale di Fisica Nucleare,
        Sezione di Ferrara, I-44100 Ferrara \\
$^{(3)}$Istituto Nazionale di Fisica Nucleare, Sezione
        di Cagliari, I-09128 Cagliari, \\ and
        Dipartimento di Fisica dell'Universit\`a di Cagliari, I-09124
        Cagliari
        }
\date{August 1995}
\maketitle                 
\begin{abstract}
We discuss what appears the last hope for an
astrophysical solution to the solar neutrino problem: a correlated
variation of the astrophysical factors for the helium burning cross
sections ($S_{33}$ and $S_{34}$) and either $S_{17}$ or the central
temperature $T_c$. In this context, we recognize the important role
played by the CNO neutrinos.
In fact, we can obtain a fair fit to the experimental data only if
three conditions are met simultaneously: the astrophysical factor
$S_{33}$ is about 200 times what is presently estimated,
the astrophysical factor $S_{17}$ is about 3 times larger and the
$^{13}$N and $^{15}$O neutrino fluxes are negligible compared to the
ones predicted by standard  solar models. These conditions are not
supported by the present data and their correlated combination is
improbable.
\end{abstract}
%
%
\narrowtext
%
\subsection{Introduction}
\label{intro}
The essence of the solar neutrino problem (SNP) is that all four
solar-neutrino experiments~\cite{Davis,Kamio,GALLEX,SAGE} detect
signals considerably smaller than the ones predicted by the standard
solar models (SSM). This deficit is  illustrated by Table~\ref{tbl1}.

There exist non standard solar models (NSSM) that predict a low
boron-neutrino flux $\Phi(\text{B})$ which is in agreement with
Kamiokande (see Table). This agreement is achieved due to a
combination of the following factors:
  the use of the data~\cite{Moto94} which indicate that the
  astrophysical factor $S_{17}$ could be smaller than the standard
  value $S_{17}=22.4$~eV barn;
  a few percent decrease of the solar central temperature caused either
  by collective plasma effects~\cite{Tsytovich} or by slightly lower
  heavy element abundances~\cite{SS93};
  a small increase of the astrophysical factor $S_{33}$ due to a
  hypothetical low-energy resonance in the reaction $^3$He + $^3$He,
  and a small decrease of the $S_{34}$ within the experimental
  uncertainties of the $^3$He + $^4$He cross section.
The  most important factor in reducing the boron flux, and consequently
in bringing NSSM in agreement with Kamiokande, is the first one,
i.e. reducing $S_{17}$.

However, the construction of such NSSM with low boron flux has not
solved the SNP, it has only shifted the emphasis from boron neutrinos
to beryllium neutrinos~\cite{GALLEX,Bahcall94b,Cast94a,Bere94a,Degl94}.
Now it is the deficit of $^7$Be
neutrinos~\cite{GALLEX,Bahcall94b,Cast94a,Bere94a,Degl94,Rosen,Rag}, or
a too low ratio of beryllium to boron neutrino flux
$\Phi(\text{Be})/\Phi(\text{B})$~\cite{Bere93a},
that constitutes the present SNP.

Several model independent analyses which use different combinations of
experimental data show clearly that the problem is real: \\
(1)The combination of the Homestake and Kamiokande data implies that,
   if neutrinos are standard (as in the SM of electroweak interactions),
   the beryllium flux has an unphysical negative value at the 92\%
   confidence level (C.L.)~\cite{BahBet,Hata,Cast93a}. This
   result is very robust.
   It does not depend either on uncertainties in nuclear reactions or
   in details of the SSM. We only need the very reliable assumption
   that the $\nu_e+{}^{37}\text{Cl} \to {}^{37}\text{Ar}+e$ cross
   section is not overestimated~\cite{Bere93a,BahBet,Hata,Cast93a}.
   Therefore, the combination of these two experiments
   strongly disfavors an astrophysical solution (with uncertainties in
   nuclear cross-sections included).\\
(2)The gallium experiments by themselves imply a deficit of the
   $^7$Be neutrinos when combined with the solar luminosity sum
   rule~\cite{gabe}: we find that now
   $\Phi(\text{Be})/\Phi^{\text{SSM}}(\text{Be})<1/2$ at the 90\% C.L.\\
(3)If we arbitrary exclude one of the four experiments from the
   analysis, e.g. we can exclude either Homestake or Kamiokande, the
   discrepancy between the remaining experiments still exists. Updating
   the analyses of Refs.~\cite{Cast94a,Degl94} leads to the limit
   $\Phi(\text{Be})< 1.5\cdot 10^9$~cm$^{-2}$~s$^{-1}$ at the
   97\% C.L.,
   which is equivalent to
    \begin{equation}
      r_{\text{Be}} \equiv
    \Phi(\text{Be})/\Phi^{\text{SSM}}(\text{Be}) < 2/7 \, ,
    \end{equation}
    where $\Phi_{\text{SSM}}(\text{Be})$ is any of the SSM fluxes
    from Table~\ref{tbl1}.\\
(4) Finally, if we use the information from all four experiments,
    we find that
    $\Phi(\text{Be})<0.7\cdot 10^9$~cm$^{-2}$~s$^{-1}$ at the 97\% C.L.
    By using any of the SSM in Table~\ref{tbl1}, this model independent
    limit can be given as
    \begin{equation}
      r_{\text{Be}} \equiv
     \Phi(\text{Be})/\Phi^{\text{SSM}}(\text{Be}) < 1/7 \, .
    \label{bebound}
    \end{equation}
    We shall use this limit in our analysis.

We have estimated the above limits and confidence levels by means of
$\chi^2$ analyses similar to those of Refs.~\cite{Cast94a,Degl94}
with the addition of Monte Carlo simulations~\cite{Hampel}
and using the new experimental data from Table~\ref{tbl1}.

It is worth observing that the restrictions discussed above  are in
fact bounds on the sum of the $^7$Be and CNO ($^{13}$N + $^{15}$O)
neutrinos. In particular,
$\Phi(\text{Be})+\Phi(\text{CNO})<0.7\cdot 10^9$~cm$^{-2}$~s$^{-1}$ at
the 97\% C.L.
The reason is that $^7$Be and CNO neutrinos have similar energies and,
therefore, similar interaction cross-sections in the detectors.
More precisely, the energy averaged
cross sections for the CNO neutrinos are slightly larger than the
cross sections for $^7$Be neutrinos. We can replace the CNO cross
sections  with that for $^7$Be neutrinos with the only consequence of
slightly underestimating the contribution of the CNO neutrinos to the
signals. Therefore,  all the above bounds apply to the sum of the
$^7$Be, $^{13}$N and $^{15}$O fluxes, since underestimating the
contribution of the CNO neutrinos only makes those inequalities
stronger.

In other words, the so-called $^7$Be neutrino problem is actually the
problem of the intermediate energy solar neutrinos.

All the above-mentioned arguments strongly suggest that
{\em non-standard neutrinos} (beyond the standard model of electroweak
interactions) are  needed to solve the SNP. In particular, we remind
that both the MSW mechanism~\cite{MSW78,MSW86} and vacuum
oscillations~\cite{Petk,rossi,calabresu} are able to explain
simultaneously all four solar-neutrino experiments.

However, as we have already briefly discussed in Ref.~\cite{Bere94b},
there is one last hope for finding an astrophysical solution to the SNP.
It consists of the following two steps:\\
(1) We strongly increase the astrophysical factor $S_{33}$, motivated by
    a hypothetical resonance in the cross section
    $^3$He + $^3$He $\to$ $^4$He + 2 p, until the $\Phi(\text{Be})$ is
    suppressed below the ``observed'' upper limit
    $\Phi^{\text{SSM}}(\text{Be}) / 7$.\\
(2) Since the first step has the undesired side effect of strongly
    suppressing also the boron flux $\Phi(\text{B})$, we boost
    $\Phi(\text{B})$ back
    to the experimental value by increasing either the astrophysical
    factor $S_{17}$ (the boron flux is directly proportional to this
    factor) or the central temperature $T_c$
    (since $\Phi(\text{B})/\Phi(\text{Be})\sim T_c^{10}$,
    $\Phi(\text{B})$ grows faster than $\Phi(\text{Be})$).\\
This game could obviously be played also with the somewhat
less stringent bounds obtained by excluding one of the experiments.
Apart from Ref.~\cite{Bere94b}, similar ideas were privately discussed
also by M.~Altman, I.~Barabanov and S.~Gershtein.

The purpose of this paper is to make a quantitative analysis of this
last hope for an astrophysical solution to the SNP. We shall study
whether it is possible to reconcile present experimental data by
simultaneously increasing  $S_{33}$ and either $S_{17}$ or $T_c$.

We shall demonstrate the important role played by CNO neutrinos,
especially for the temperature solution.

\subsection{Analytical approach}
\label{ana}
For the following semiquantitative analysis,
we use two scaling laws~\cite{Cast94b}:
\begin{equation}
\Phi(\text{Be}) \sim T_c^{9}\, S_{34}\, S_{33}^{-1/2}
\label{belaw}
\end{equation}
and
\begin{equation}
\Phi(\text{B}) \sim T_c^{22}\, S_{17}\, S_{34}\, S_{33}^{-1/2} \, .
\label{blaw}
\end{equation}
Had we used instead the not too different scaling laws of
Ref.~\cite{Bahcall89}, results would have been  similar.
Moreover, for simplicity, we only discuss the roles of the boron and
beryllium fluxes; however, we shall eventually comment on the relevance
of the other fluxes, especially those of the CNO cycle.

The fact that the beryllium and boron fluxes
depend on $S_{33}$ and  $S_{34}$ only through the ratio~\cite{Cast93a}
\begin{equation}
X\equiv S_{33}/S_{34}^2
\end{equation}
implies that our analysis automatically includes not only an
increase of $S_{33}$ but also a decrease of $S_{34}$.
In the following we shall use the notation
\begin{eqnarray}
\label{xs33}
x&\equiv& X/X^{\text{SSM}} = \frac{S_{33}}{S^{\text{SSM}}_{33}}/
                \left(\frac{S_{34}}{S^{\text{SSM}}_{34}}\right)^2 \\
s_{17}&\equiv& S_{17}/S^{\text{SSM}}_{17}\\
t_{c}&\equiv& T_{c}/T_c^{\text{SSM}}\, .
\end{eqnarray}

First, let us analyze the solution that involves only adjusting the
nuclear cross sections and assumes $T_c=T_c^{\text{SSM}}$, i.e. $t_c=1$.
Using the scaling law of Eq.~(\ref{belaw}) one obtains from
the bound on beryllium flux, Eq.~(\ref{bebound}):
\begin{equation}
r_{\text{Be}} \equiv \frac{\Phi(Be)}{\Phi^{SSM}(Be)}=
\frac{1}{\sqrt{x}} \leq 1/7 \, ,
\end{equation}
which results in
\begin{equation}
x \geq 50 \, .
\label{xbound}
\end{equation}
{}From the scaling law of Eq.~(\ref{blaw}), using
the one-sigma lower limit for the $^8$B flux from
the Kamiokande experiment,
$\Phi(\text{B}) \geq 2.3 \cdot 10^{6}$~cm$^{-2}$~s$^{-1}$, and
the SSM value $\Phi_{\text{SSM}}(\text{B})=6.62 \cdot
10^{6}$~cm$^{-2}$~s$^{-1}$ we find:
\begin{equation}
r_{\text{B}} \equiv \frac{\Phi(\text{B})}{\Phi_{\text{SSM}}(\text{B})}=
\frac{s_{17}}{\sqrt{x}}
\leq 0.35 \,
\end{equation}
Combined with the bound on $x$, given by Eq.~(\ref{xbound}), it yields
\begin{equation}
s_{17} > 2.4\, .
\end{equation}
If we use Bahcall scaling relations~\cite{Bahcall89},
instead of Eqs~(\ref{belaw}) and~(\ref{blaw}), we obtain the stronger
limits
\begin{eqnarray}
x &\geq& 130\\
s_{17} &>& 2.4\,.
\end{eqnarray}

Now, let us examine the ``temperature solution'', assuming
$S_{17}=S_{17}^{SSM}$ and increasing $T_c$.
Since in this case it is not possible to derive a strict inequality,
we use the central value for the Kamiokande result
$\Phi(\text{B}) = 2.73 \cdot 10^{6}$~cm$^{-2}$~s$^{-1}$, and the
upper limit for
the beryllium flux (a more complete numerical analysis that takes
properly into account  uncertainties, other fluxes, etc. can be found
in the next section).
With these fluxes we obtain
\begin{equation}
r_{\text{Be}}=\frac{t_c^9}{\sqrt{x}} \approx 1/7
\end{equation}
and
\begin{equation}
r_{\text{B}}=\frac{t_c^{22}}{\sqrt{x}} \approx 0.41 \, ,
\end{equation}
and thus
\begin{eqnarray}
t_c &\approx& 1.08\\
x& \approx& 210 \, .
\end{eqnarray}

We can conclude as follows:\\
(1) In both cases,  an extremely high value of $x$, and therefore of
$S_{33}$, is necessary. It clearly implies a so far undetected,
and theoretically disfavored,
very-low-energy (say $E_r<20$~KeV) resonance in the $^3$He + $^3$He
channel. Nonetheless, this possibility cannot be completely ruled out
and it is presently being investigated in an experiment at the LNGS
underground laboratory~\cite{Arpe91}.\\
(2) The first case ($x$ and $s_{17}$ increase), in addition to a large
value of $S_{33}$, requires also a value of $S_{17}$ almost three times
larger than the one used in the SSM:
$S_{17}^{\text{SSM}}=22.4$~eV~barn. Should we accept the recently
proposed smaller value~\cite{Moto94}, the situation would be hopeless.\\
(3) In the second case ($x$ and $t_c$ increase),
in addition to very large value of $S_{33}$, an increase of the
central temperature by 8\% is also needed. This value is difficult to
reconcile with the helioseismological data.
In addition, and more important, if one tries to enhance the solar
temperature, the CNO cycle gains  efficiency, and the production of
$^{13}$N and $^{15}$O neutrinos grows as fast as the one of $^{8}$B
neutrinos. This effect will be analyzed numerically in the next
section.\\
(4) Moreover, in the SSM, the CNO neutrinos alone already saturate the
bound $\Phi(Be)+\Phi(CNO)<0.7\cdot 10^9$~cm$^{-2}$~s$^{-1}$.
Even if we were able to suppress $^7$Be neutrinos almost to zero
and, at the same time, we could make the
boron flux compatible with experiments,  there will be still a conflict
with experiments  due to the CNO neutrinos. This is the crucial point
for understanding why both attempts outlined above fail even more
miserably when we  perform the numerical calculations taking into
account the CNO neutrinos.

\subsection{Numerical analysis}
\label{num}
In this section we verify numerically the conclusions reached by
the semiquantitative analytical analysis in the previous section.
Here we include in the calculations the CNO neutrinos,
treat  more accurately the dependence on $S_{33}$, $S_{17}$
and $T_c$ and take into account uncertainties in the input parameters.
Regarding the dependence of
the boron and beryllium flux on $T_c$,
it is worth observing  that a low energy $^3 He + {}^3$He resonance
suppresses beryllium neutrinos  more
than boron neutrinos:  $r_{\text{Be}}=0.76
\cdot r_{\text{B}}$~\cite{Cast93a}.
The reason is  that this low energy resonance
is more effective at lower temperatures, and thus it is more efficient
in the outer (cooler) region  where $^7$Be neutrinos are produced.
This effect is taken into account in the numerical calculations, which,
for simplicity, we still present in terms of an effective
$s_{33}=S_{33}/S_{33}^{\text{SSM}}$.

Without loss of generality and for the sake of two-dimension graphical
presentation we shall use $S_{34}=const$ . If one is interested in the
explicit dependence on $S_{34}$, he can replace everywhere
$s_{33}$ with $x$ given by Eq.~(\ref{xs33})
as discussed in the previous section.

\subsubsection{The correlated variations of $S_{33}$ and $S_{17}$}
The numerical results confirm the analytical estimates: extremely
large values of $S_{33}$ and $S_{17}$ are needed. They are larger
than in the analytical estimate mainly due to the already mentioned
contribution of the  CNO neutrinos.  As it is clearly seen in
Fig.~\ref{fig1}, even for values of $s_{33}\approx 200$ and
$s_{17}=3.4$, the $\chi^2$ is still above 15 (at the 99\% C.L. the
$\chi^2$ for 4 degrees of freedom should be less than 13.28).

This result can be understood from Fig.~\ref{fig2}. The solid
line going from the diamond labeled SSM to the point  1
shows the effect of increasing $s_{33}$ by a factor 170: the final value
of $\Phi(\text{Be})+\Phi(\text{CNO})$
is still twice the value allowed at the 95\% C.L. by the experiments
and it is therefore useless to try to adjust the boron
flux by increasing $S_{17}$ (solid line between points 1 and 2).

The same figure shows that the main problem is due to neutrinos from the
CNO cycle (consider the dashed lines):
if we arbitrary switch off the $^{13}$N and $^{15}$O fluxes (point
labeled NO CNO) our game of taking $s_{33}=170$ (point 4) and then
$s_{17}=3$ produces the  point 5, which shows that it is possible to
reach the region allowed at the 95\% C.L. (solid ellipse). However,
even in this unrealistic case the values of $S_{33}$ and/or $S_{34}$
are too high.

\subsubsection{The correlated variations of $S_{33}$ and $T_{c}$}
The numerical results here are even more discouraging than in the
previous case (see Fig.~\ref{fig3}). The minimum  $\chi^2$ is this
time larger than 20. Moreover, it is clear that increasing the
temperature does not help, and the ``best'' results are actually
obtained for reduced values of the temperature.

Figure~\ref{fig2} can again help us to understand the reason of such
behavior. As before taking $s_{33}=170$ (point 1) does not reduce
sufficiently the sum of the beryllium and CNO fluxes; in addition, if we
increase $T_{c}$ the CNO fluxes increase as fast as the boron flux with
the result that the point 3 is far away from the allowed region (solid
ellipse).

As in the previous case, we can stress the importance of the CNO
neutrinos by considering the same solar model with their contribution
reduced to zero. Now increasing the temperature
we are able to barely reach the 95\% C.L. allowed region
(see dashed line from point 4 to point 6).

\subsubsection{Eliminate one experiment?}
Given the well-known ``incompatibility'' of the experimental results,
one might think that disregarding one of the experimental result could
be the solution to our problems. We find that the situation does not
change drastically. In particular, the most favorable case, which
corresponds to neglecting the Kamiokande result and to a variation of
$s_{33}$ and  $s_{17}$, still gives us a $\chi^2$ greater than 8
(at the 95\% C.L. the $\chi^2$ for 3 degrees of freedom should be less
than 8) for $s_{33}=200$ and $s_{17}=2.5$.
The basic reason is that even if we have eliminated the Kamiokande
constraint on the boron flux, this flux cannot be much smaller than
before, as it can be seen comparing the two 95\% C.L. regions in
Fig.~\ref{fig2}: the one obtained using all four experiments (solid
ellipse) and the other obtained using only the chlorine and gallium
data (dotted ellipse). In fact the chlorine result implies that the
contribution from beryllium and CNO neutrinos must increase if the
boron flux is too low, but the gallium result forbids a too high
beryllium flux. Therefore, we are still only able to get close to the
allowed region, if the CNO neutrino fluxes are not much smaller than
the ones predicted in SSM (see solid line from point 1 to point 2).
\subsection{Conclusions}
\label{conclu}
We have discussed what appeared to be the last hope for an
astrophysical solution to the SNP, i.e. a correlated variation of
$S_{33}/S_{34}$ and either $S_{17}$ or the central temperature $T_c$.
The important role played by the  CNO neutrinos has been properly
emphasized in our discussion.

We have concluded that:\\
(1)If  the calculated fluxes of the CNO neutrinos ($^{13}$N and
   $^{15}$O) are not greatly overestimated, there is absolutely no
   chance of  solving the SNP by adjusting  $S_{33}$ and/or $S_{34}$,
   and either $S_{17}$ or the central temperature $T_c$.\\
(2)Even if the CNO fluxes were negligible and a hypothetical low energy
   resonance allowed us to increase $S_{33}$ at our convenience, we
   would still need an astrophysical factor $S_{17}$ about 3 times
   larger than the SSM  value. There is no experimental indication of
   such an enhancement; on the contrary, it is claimed~\cite{Moto94}
   that the actual value is even smaller.\\
(3)The situation is even worse if one tries to increase the solar
   temperature. Sooner or later the CNO cycle becomes  efficient and one
   is again producing too many intermediate energy neutrinos
   (see Fig.~\ref{fig2}).\\
(4)If one arbitrarily disregards any single experiment, we still need a
   strong reduction of the CNO fluxes and a large increase (close to
   200) of $S_{33}$.

Thus, the last hope turned out to be a no-hope case.

\begin{table}
\caption[xman]{Comparison of the most recent
    experimental data (Experiment), and a selected sample of
    theoretical predictions including some from low-boron-neutrino-flux
    models (SS93 and DS94). We also report predictions
    for the main fluxes (pp, $^7$Be, $^{13}$N, $^{15}$O and $^8$B) and
    for the central temperature $T_c$, and the model input values for
    the astrophysical factors $S_{33}$, $S_{34}$ and $S_{17}$. Only the
    sum $^{13}$N + $^{15}$O is reported for SS93. For the experimental
    data we give separately statistical and systematic errors, while
    for the theoretical predictions errors are 1$\sigma$ ``effective''
    errors.
         \label{tbl1}
         }
\begin{tabular}{ccccccc}
 &\multicolumn{3}{c}{Standard models}
             &\multicolumn{2}{c}{Low-flux models} &\\
                & TCL93~\tablenote{Ref.~\protect\cite{TCL93}}
                & CDF94~\tablenote{Ref.~\protect\cite{Cast94b}}
                & BP95~\tablenote{Ref.~\protect\cite{BP95}}
                & SS93~\tablenote{Ref.~\protect\cite{SS93}}
                & DS94~\tablenote{Ref.~\protect\cite{DS94}}
                & Experiment\\
\tableline
 pp             & 60.2    & 60.0    & 59.1     & 61     & 60.4  & \\
{}[10$^9$ cm$^{-2}$ s$^{-1}$] & & & & & & \\
$^7$Be          & 4.33    & 4.79      & 5.15   & 3.9    & 4.30    & \\
{}[10$^9$ cm$^{-2}$ s$^{-1}$] & & & & & & \\
$^{13}$N        & 0.382   & 0.47      & 0.618  &        & 0.075     & \\
{}[10$^9$ cm$^{-2}$ s$^{-1}$] & & & & 0.3 & & \\
$^{15}$O        & 0.318   & 0.40      & 0.545  &        & 0.022     & \\
{}[10$^9$ cm$^{-2}$ s$^{-1}$] & & & & & & \\
$^8$B/Kamiokande & $4.4\pm 1$  & 5.6       & $6.6^{+0.9}_{-1.1}$
                                      & 3.0     & 2.77     &
$2.75^{+0.20}_{-0.18}\pm 0.41$~\tablenote{Ref.~\protect\cite{Kamio}} \\
{}[10$^6$ cm$^{-2}$ s$^{-1}$] & & & & & & \\
GALLEX          & $ 122\pm 7$ & $130\pm 7$ & $ 137^{+8}_{-7}$ & 114
       & 109 &
$77.1\pm 8.5^{+4.4}_{-5.4}$~\tablenote{Ref.~\protect\cite{GALLEX}}\\
{}[SNU]           & & & & & & \\
SAGE       & $ 122\pm 7$ & $130\pm 7$ & $ 137^{+8}_{-7}$ & 114 & 109
      & $69\pm 11\pm 6$~\tablenote{Ref.~\protect\cite{SAGE}}\\
{}[SNU]           & & & & & & \\
Chlorine        & $6.36\pm 1.3$ & $7.8\pm 1.4$   &
 $9.3^{+1.2}_{-1.4}$ & 4.5 & 4.2 &
       $2.55\pm 0.17 \pm 0.18$~\tablenote{Ref.~\protect\cite{Davis}}\\
{}[SNU]           & & & & & & \\
$S_{33}$        & 5.24 & 5.00 & 4.99 & 5.6 & 5.6 & \\
{}[MeV barn]      & & & & & & \\
$S_{34}$        &      & 0.533 & 0.524 &  & 0.45 & \\
{}[KeV barn]      & & & & & & \\
$S_{17}$        & 22.4 & 22.4 & 22.4  & 20.2 & 17 & \\
{}[eV barn]       & & & & & & \\
$T_c$           & 1.543 & 1.564 & 1.584 & 1.545 & 1.571 & \\
{}[$10^7$ K]      & & & & & & \\
\end{tabular}
\end{table}
\begin{figure}
\caption[s33s17]{
   Contours of equal $\chi^2$ for the neutrino fluxes in nonstandard
   solar models parameterized by the $S_{33}$ and $S_{17}$
   astrophysical factors, which have been normalized to their SSM
   values (5.00 [MeV barn] and 22.4 [eV barn], respectively). Solid
   contours correspond to $\chi^2$ equal to 40, 35, 30, 25 and 20;
   broken contours correspond to values in between. Note that values
   of $\chi^2 > 13.28$ have less than  1\% probability for the four
   data (chlorine, GALLEX, SAGE and Kamiokande).
               }
\label{fig1}
\end{figure}
\begin{figure}
\caption[pippino]{
Beryllium plus CNO fluxes vs. boron flux.
The solid ellipse confines the region allowed at the 95\% C.L. by the
four current experiments (chlorine, GALLEX, SAGE and Kamiokande)
with the only constraint due to the luminosity sum rule.
The dotted ellipse confines the region allowed by only the chlorine
and gallium experiments. The diamond shows the SSM prediction.  When
increasing $S_{33}$ (or more generally $x$), the theoretical point
moves along the solid line and reaches point 1 at $x=170$. If, starting
from this point 1, we increase $t_c \equiv T_c/T_{SSM}$, the point
moves away from the allowed region towards point 3 and reaches it at
$t_c=1.07$. This unsuccessful game with the "temperature solution" is
caused by the increase of the CNO flux with the temperature. If instead,
starting again from point 1, we increase $s_{17}$, the theoretical
point moves towards point 2 and reaches it at $s_{17}=3$. This point is
still outside the allowed region. The same SSM, but with the $^{13}$N
and $^{15}O$ fluxes reduced to zero, is labeled NO CNO. Points 4, 5, 6
are the analogues of points 1, 2, 3, respectively. The ``NO CNO'' track
clearly illustrates the role of CNO neutrinos. However, even for this
track (absence of CNO neutrinos), the theoretical point gets into
allowed region at too large values of $S_{33}$ and $S_{17}$.
               }
\label{fig2}
\end{figure}
\begin{figure}
\caption[s33tc]{
   Contours of equal $\chi^2$ for the neutrino fluxes in nonstandard
   solar models parameterized
   by the $S_{33}$ astrophysical factors and the central temperature
   $T_c$, both of which have been normalized to
   their SSM values (5.00 [MeV barn] and $1.564\times 10^7$~[K],
   respectively).
   Solid contours correspond
   to $\chi^2$ equal to 40, 35, 30 and 25; broken contours correspond
   to values in between.  Note that values of $\chi^2 > 13.28$
   have less than 1\% probability for the four data (chlorine, GALLEX,
   SAGE and Kamiokande).
               }
\label{fig3}
\end{figure}
\end{document}